# Finite Element Prediction of Sintering Deformation in 3D-Printed Porcelain Filament

Fatima Hammoud[a*] and Charles Manière[a*]

[a] Université de Caen Normandie, ENSICAEN, CNRS, CRISMAT, Normandie Univ, 14000 Caen, France

[*] Corresponding authors: fatima.hammoud@ensicaen.fr, charles.maniere@ensicaen.fr

**Abstract.** Sintering of printed porcelain filaments can be strongly affected by overhang geometry, thin features, and printing-induced anisotropy. These effects are particularly difficult to simulate because they require accurately capturing the interplay between sintering kinetics, viscous deformation, and macroscopic anisotropy. In this study, a robust and predictive model is established experimentally using anisotropic sintering dilatometry combined with overhang-deformation calibration through finite element simulations (FEM). The sintering behavior is identified using a porosity-temperature-independent minimization strategy applied to multi-rate dilatometry. Pellet anisotropy is incorporated via FEM by imposing directional sintering stresses. The deformation behavior is then calibrated by adjusting the shear viscosity to match overhang shapes deformation. The resulting model is finally validated on overhanging bar geometries. This step-by-step approach provides a comprehensive assessment of porcelain sintering while accurately capturing deformation mechanisms.

**Nomenclature**

$\theta$ Porosity

$\dot{\theta}$ Porosity elimination rate (s$^{-1}$)

$\rho$ Relative density

$\underline{\sigma}$ Stress tensor (N.m$^{-2}$)

$\underline{\dot{\varepsilon}}$ Strain rate tensor (s$^{-1}$)

$\eta$ Material viscosity (Pa.s)

$\eta_0$ Viscosity pre-exponential factor (Pa.s)

$Q$ Deformability activation energy (J.mol$^{-1}$)

$R$ Gas constant 8.314 (J.mol$^{-1}$.K$^{-1}$)

$T$ Temperature (K)

$\varphi$ Shear modulus

$\psi$ Bulk modulus

$\theta_{ci}$ Initial critical porosity

$\theta_{cf}$ Final critical porosity

$P_l$ Sintering stress (N.m$^{-2}$)

$P_{lR}$, $P_{lZ}$ Anisotropic sintering stress terms (N.m$^{-2}$)

$\bar{\bar{I}}$ Identity tensor

$\alpha$ Surface energy (J.m$^{-2}$)

$r$ Particles radius (m)

$a, \gamma, \zeta, \beta, \chi$ Fitting constants

$t$ Time (s)

$\Phi$ Master sintering curve integral parameter

MSC Master Sintering Curve

## 1. Introduction

Porcelain is a well-established class of traditional ceramics, extensively used in applications ranging from tableware and decorative items to advanced uses such as dental restorations, insulators, and technical substrates [1] [2] [3] [4] [5]. Its widespread use is largely due to its desirable properties, including high mechanical strength, low water absorption, thermal and chemical stability, and aesthetic translucency [6] [7]. Technologically, porcelain is composed of a ternary mixture of kaolin (clay), feldspar, and quartz. During firing, feldspar acts as a fluxing agent by forming a viscous liquid phase that facilitates densification and promotes the formation of mullite, the primary crystalline phase responsible for mechanical integrity and thermal performance [8] [9]. The final microstructure typically contains mullite crystals embedded in a glassy matrix, with residual undissolved quartz grains.

Sintering plays a crucial role in defining the final properties of porcelain, as it governs densification, phase evolution, and dimensional stability. The kinetics of sintering are strongly influenced by temperature, dwell time, heating rate, and chemical composition [10] [11]. Mullite formation initiates at temperatures above 1100 °C and generally stabilizes between 1200-1400 °C [12] [13], while the composition of the glassy phase depends on the type of feldspar used (e.g., potash or soda feldspar) and its eutectic behavior [7]. Moreover, quartz dissolution becomes increasingly significant above 1200 °C, contributing to the silica network within the glassy phase and influencing shrinkage behavior and mechanical properties [14].

Due to the high thermal demands of conventional sintering, which often involves slow heating rates and long processing cycles, significant research has focused on reducing energy consumption without compromising product quality [15] [16]. Innovations such as fast firing, direct sintering (DS) [17], and field-assisted sintering techniques (FAST) have been explored as alternatives to reduce processing time and peak temperatures [18] [19]. While these approaches can be effective, they often require specialized equipment or strict process control, and their suitability for more complex geometry parts remains a challenging due to scalability issues and high distortions [20]. Since the sintering process involves a liquid phase, long conventional sintering cycles are particularly prone to undesired distortions, which pose a challenge ranging from large sanitaryware applications to delicate, thin, overhanging artistic structures [21] [22] [23].

The recent introduction of Zetamix by Nanoe® filaments offers a promising route for shaping porcelain components via fused filament fabrication (FFF), a form of extrusion-based additive manufacturing [24] [25]. Recent AM studies on polymers and composites, such as PLA–ZnO nanocomposites [26] and hydrogel filaments [27], focus mainly on pre-sintering behavior. In contrast, this work examines the viscous sintering of Zetamix® Nanoe porcelain by coupling MSC/WR kinetics with finite element simulations validated on printed geometries. These filaments consist of a high ceramic load in a thermoplastic binder, enabling precise, layer-by-layer deposition of green bodies prior to thermal debinding and sintering [24] [28]. This method provides flexibility in design and geometry, reduces material waste, and opens new opportunities for customized or functional porcelain parts [29] [30]. However, the layered structure and binder composition introduce new challenges during sintering, such as anisotropic shrinkage, non-uniform densification, and warping, which can compromise dimensional accuracy and mechanical performance [31] [28] [32].

In this context, predictive modeling of sintering behavior becomes essential. Finite element modeling (FEM) allows for the simulation of complex thermo-mechanical phenomena, including densification, viscous flow, and anisotropic shrinkage, especially when calibrated with reliable experimental data [33] [34] [35]. By integrating modeling with physical measurements, such as slumping and deflection tests to determine shear viscosity, one can gain deeper insights into the evolution of strain and porosity during sintering.

Beyond polymeric systems, recent post-AM sintering studies on metals, such as stainless steel [36] and extrusion-based alloys [37], focus on solid-state mechanisms. In contrast, this work examines viscous-flow sintering in a multiphase porcelain with a transient liquid phase, using a temperature-dependent viscosity and directional sintering stress to capture anisotropic shrinkage. The model is carefully calibrated and validated through dilatometry and deformation tests on printed geometries.

The objective of this study is to investigate the anisotropic sintering behavior of porcelain parts produced using Zetamix by Nanoe® filaments under different heating rates (2, 5, and 10 K/min), and to develop a coupled numerical-experimental methodology to model shrinkage and densification. Finite element simulations are conducted to capture axial and radial displacements, porosity evolution, and strain development. The model is calibrated using experimental data from thin-cylinder slumping and bar deflection tests. Furthermore, the Master Sintering Curve (MSC) [38] approach and Wang-Raj (WR) [39] method are applied to evaluate sintering activation energy and validate the thermally activated nature of densification. Then, a multivariable adjustment method is used to determine the sintering moduli [40], and the observed sintering-induced warping is used to refine the shear modulus that governs part deformation at high temperatures [23]. This integrated approach provides a framework for optimizing the sintering process of 3D-printed porcelain and contributes to the advancement of energy-efficient, geometry-accurate ceramic manufacturing.

## 2. Materials and Methods

*2.1. Samples Preparation*

Zetamix by nanoe® porcelain filaments, specifically formulated for ceramic fused filament fabrication (FFF), were used in this study. These filaments are composed of fine porcelain powder dispersed in a polyolefin-based thermoplastic binder. Due to their relatively soft nature, care was taken during filament loading to avoid grinding or deformation. A grooved driving gear was used to ensure smooth feeding, and the extruder was preheated prior to printing to allow material extrusion and detect possible nozzle obstructions.

Printing was performed on a Prusa i3 MK3S printer using a 0.6 mm brass nozzle and PrusaSlicer 2.5.0. The parameters were set according to manufacturer recommendations: a nozzle temperature of 140 °C, bed temperature of 50 °C, layer thickness of 0.2 mm, and printing speed between 20 and 40 mm/s. Additional settings included 3 wall lines, full fan speed (100%), and optional retraction. Grid infill patterns were used with 100% densities to print 8 mm diameter, 3 mm height pellet sample, bar and circle specimens for FEM model validation. Top and bottom surfaces were defined by four layers.

After printing, the green parts underwent a two-stage debinding process. The first step involved solvent debinding in an acetone bath maintained at 40 °C for at least 2 hours, depending on the geometry and wall thickness. This step aimed to remove about 12 wt% of the binder. After solvent treatment, the parts were air-dried at room temperature for 12 hours.

The second step was thermal debinding, conducted in ambient air. The heating profile involved a slow ramp from 50 °C to 500 °C at 7 °C/h, followed by a 2-hour dwell. The complete cycle lasted approximately 2.5 days.

*2.2. Sintering process*

Sintering of the debound porcelain parts was carried out in a high-temperature furnace under ambient atmosphere. The temperature cycle involved a ramp-up from 50 °C to 1250 °C at a constant rate of 120 °C/h, followed by a holding period of 2 hours at 1250 °C to promote densification. After the dwell time, the samples were cooled down to 50 °C at the same rate, resulting in a total cycle duration of approximately 22 hours.

The selected parameters followed the manufacturer's recommended sintering profile for Zetamix® porcelain filaments and were validated through preliminary dilatometry and microstructural analyses, confirming optimal densification without excessive warping or cracking.

The sintering process was designed to achieve full densification while managing dimensional changes. Based on Zetamix specifications and experimental measurements, the linear shrinkage observed after sintering was approximately 13.8% ± 1% in the X and Y directions, and 26.4% ± 1% in the Z direction. These values were accounted for in the design phase by applying corresponding oversizing factors during slicing.

*2.3. Characterization*

The thermal and densification behavior of the porcelain samples was characterized using dilatometric analysis at three different heating rates (2, 5, and 10 K/min) with 2 h of holding at 1300°C to ensure optimal densification for the model data. Linear shrinkage in the radial and vertical directions (building direction) was measured as a function of time and temperature to account the inherent sintering anisotropy. Relative density ($\rho$) was estimated from final relative density ($\rho_f$) the height (H) and radius (R) of the pellet with the following equation.

$$\rho_t = \frac{\rho_f H_f R_f^2}{H_t R_t^2} \tag{1}$$

In addition, optical microscopy was used to observe the microstructure of sintered cross-sections. These observations revealed the final porosity of the microstructures as a function of heating rate, providing qualitative confirmation of the densification trends obtained by dilatometry.

Thermal analysis results were used to construct sintering curves, which were processed using the Master Sintering Curve (MSC) and Wang-Raj (WR) methods and the "Sinterlab" software. These models were applied to extract activation energy values and identify the thermally activated nature of the densification process.

Material parameters for sintering modeling were identified using the following regression equation and using the modified Skorohod moduli [40]. The optimal values of the exponents were determined by minimizing the difference between activation energies obtained from MSC and the regression.

$$Y = ln\left(\frac{-3(1-\theta)^3}{T\dot{\theta}\psi}\right) = ln\left(\frac{r\eta_0}{\alpha}\right) + \frac{Q}{RT} \tag{2}$$

$$\psi = \frac{2}{3}\frac{(\theta_{ci}-\theta)^\gamma}{(\theta-\theta_{cf})^\zeta} \tag{3}$$

Model validation was conducted by comparing simulated and experimental results for porosity ($\Theta$) evolution and shrinkage in both the radial and vertical directions. An analytic modeling approach was used to plot the predicted porosity evolution with the following equation. The anisotropy is added to the model directly in a first FEM simulation of the pellet sintering in the dilatometer. FEM simulations were then performed on realistic geometries, including slumping of thin cylinders and deflection of sintered bars. The resulting displacement fields were compared to experimental measurements for model verification.

$$\theta_{t+1} = \theta_t + \Delta t \; \frac{-3(1-\theta)^3}{\frac{r\eta_0}{\alpha} T \exp(\frac{Q}{RT}) \psi} \tag{4}$$

*2.4. Finite Element Modelling*

Finite element simulations were carried out to model the anisotropic shrinkage, densification, and distortion during the sintering of porcelain parts. The simulation was performed using COMSOL Multiphysics® software. The sintering behavior is modeled by the following equation that relates the stress and strain rate tensors.

$$\underline{\sigma} = 2\eta(T)\left(\varphi(\theta)\underline{\dot{\varepsilon}} + \left(\psi(\theta) - \frac{1}{3}\varphi(\theta)\right)\dot{e}\mathbb{1}\right) + P_l \mathbb{1} \tag{5}$$

For porcelain, the viscosity expression is not grain size dependent like for solid state diffusion and have the following expression.

$$2\eta = \eta_0 T \exp(\frac{Q}{RT}) \tag{6}$$

With the following Skorohod's expression of the sintering stress [41].

$$P_l = \frac{3\alpha}{r}(1-\theta)^2 \tag{7}$$

The masse conservation is then used to actualize the densification which the trace of the strain rate tensor.

$$\frac{\dot\theta}{1-\theta} = \dot{\varepsilon}_x + \dot{\varepsilon}_y + \dot{\varepsilon}_z \tag{8}$$

The model focuses on the sintering stage following debinding. Residual stresses and microcracks potentially generated during the debinding step are assumed to relax before sintering and are thus not explicitly modeled. Their macroscopic effect is implicitly captured in the calibrated viscosity and anisotropic stress parameters.

The geometries investigated in this study (pellets, slumped rings, and deflected bars) were deliberately selected to isolate specific sintering mechanisms under controlled boundary conditions: anisotropic densification, gravity-driven slumping/bending. This systematic validation ensures that each deformation mode is accurately represented before extending the model to more complex FFF-produced geometries, such as overhanging or multi-thickness parts, which motivated the development of this predictive framework.

The boundary conditions consist of prescribed displacement edge/points that stabilize the object on the alumina support edge without overconstraining it. This assumption is justified by the selected ring shape and the much smaller gap between the porcelain part and the alumina support.

*2.5. Sintering model identification by thermally/porosity-independent assessment method*

Sintering curves were derived from dilatometry measurements performed at various heating rates. These curves were used to define the temporal evolution of relative density in the numerical model. Time, temperature integration was carried out using the sintering integral, allowing the representation of porosity independent thermally activated densification by MSC and Wang and Raj approaches. Then, regression minimization was conducted using (2) to assess the porosity dependent bulk modulus exponents (3). The densification model is then tested using (4). Next, the FEM simulation of the dilatometry pellet was conducted to adjust the anisotropic sintering shrinkage by anisotropic sintering stress ($Pl_R$, $Pl_Z$). Then the last remaing parameter to adjust is the material deformation by creep at high temperature. A thin circular cylinder is sintered on the side to detect a deformation. The FEM simulation is then conducted to reproduce this deformation. The following shear viscosity term ($G$) is adjusted so the simulation corresponds to the observed deformation.

$$G = \eta\varphi = \eta\beta(1-\theta)^\chi \tag{9}$$

Where $\beta$ and $\chi$ are adjustable coefficients of the shear viscosity.

The main stage of the study are summarized in Figure 1.

| Sintering Dilatometry | Sintering Curves (input model data) | Model identification |
|---|---|---|

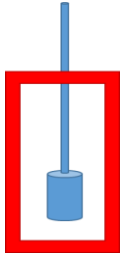 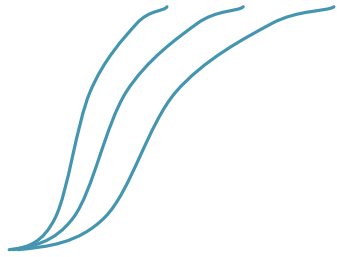 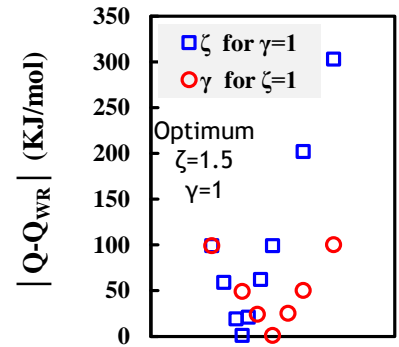

$$\psi = \frac{2}{3}\frac{(\theta_{ci}-\theta)^{\gamma}}{(\theta-\theta_{cf})^{\zeta}}$$

| FEM simulation of distorsion | Dilatometry model validation |
|---|---|

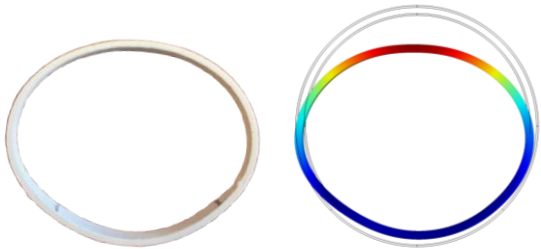 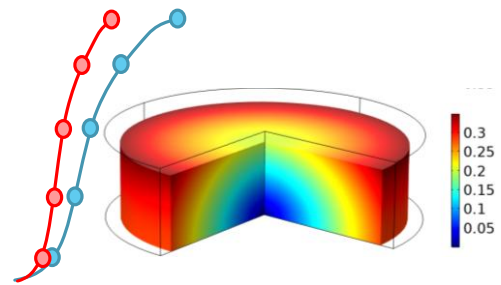

*Figure 1. Overview of the modeling process combining dilatometry data, sintering curve fitting, parameter identification, and FEM simulation, followed by experimental validation.*

## 3. Results and discussions

*3.1. Shrinkage and Densification Behavior*

Understanding how porcelain parts shrink and densify during sintering is essential for predicting final geometry and guiding process optimization. To this end, we first evaluated the thermo-mechanical response of printed samples using dilatometry at different heating rates.

Figure 2 presents the linear shrinkage (radial and vertical) and relative density evolution for samples sintered at 2, 5, and 10 K/min. At lower heating rates, shrinkage initiates gradually and progresses over a broader temperature range. In contrast, sintering at 10 K/min results in a steeper shrinkage curve, indicating a more rapid densification process that quickly attains a threshold near 97%.

Relative density increases from approximately 70% in the green state to nearly 98% after sintering. A change in the densification regime is detected near 1000°C for all curves. This behavior is well known and corresponds the appearance of liquid phase, rearrangement, and the onset of mullite needle crystallization (primary mullite) [42]. This phenomenon is detectable but occurs during the early stage of sintering, where only a small amount of liquid phase is present. Therefore, it will be neglected in the simulation, as it does not significantly contribute to deformation, unlike final-stage sintering. It is also important to note the decrease in sintering shrinkage anisotropy with the increase in heating rate. For the 10 K/min test, the anisotropy is inverted. This suggests that low heating rates give more time for mullite particle rearrangement in the liquid phase, which results in exaggerated anisotropy [43]. The simulation will focus on the 2 K/min test, as it corresponds to the industrial cycle.

Although the dilatometry curves reveal anisotropic shrinkage, the Master Sintering Curve (MSC) method is later applied only to determine the apparent activation energy of densification under controlled stress-free conditions. The anisotropic and gravity-induced effects are instead handled within the finite element model, which explicitly incorporates direction-dependent sintering stresses.

The dilatometry tests at 5 and 10 K/min were used to assess the thermally activated nature of densification (activation energy) by porosity independent kinetic-based analysis through the MSC and Wang-Raj analyses. Although the 2 K/min condition was chosen for FEM calibration to represent the industrial process, the higher heating rate data were essential to ensure the validity and transferability of the sintering kinetics.

It should be noted that the industrial sintering cycle reaches approximately 1250 °C, whereas the dilatometry tests were extended to 1300 °C to capture the complete densification curve and identify the plateau region. The model calibration was restricted to the temperature range below 1250 °C, consistent with the actual process conditions.

The experimental curves obtained here serve as a foundation for calibrating the sintering model. They provide the time and temperature resolved data required to construct sintering kinetics and extract activation parameters in the following sections.

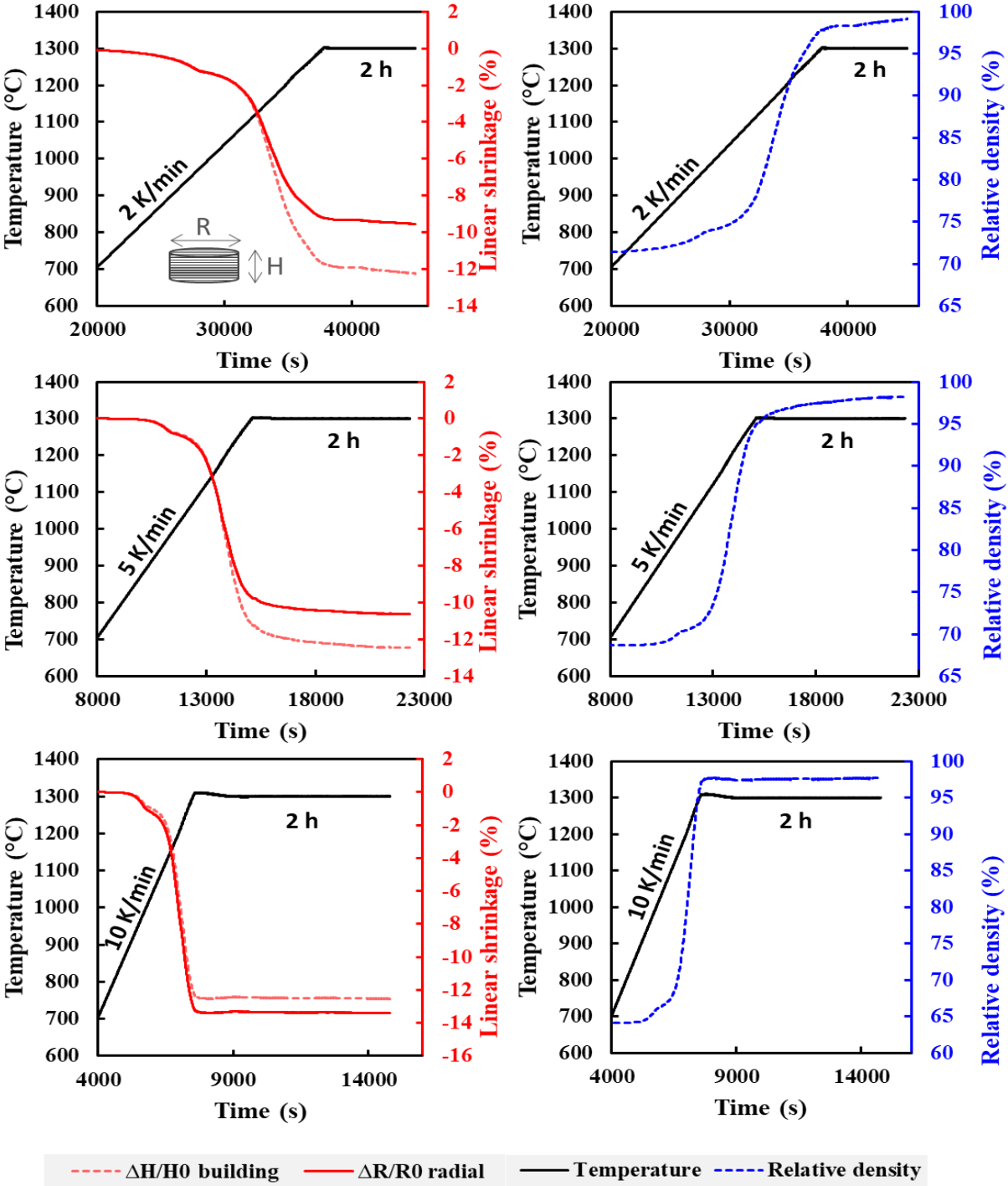

*Figure 2. Dilatometric curves showing radial and vertical shrinkage, temperature, and relative density during sintering of porcelain at different heating rates.*

To further illustrate the effect of heating rate on microstructural evolution, optical microscopy was performed on polished cross-sections of sintered samples (Figure 3). The micrographs reveal a dense matrix containing nearly circular pores and darker silica-rich regions. The circular pore morphology suggests isotropic surface-tension-driven coalescence during the final sintering stage where liquid phase is present. The silica-rich regions are very typical of porcelain where silica is used as reinforcement charge. Increasing the heating rate slightly reduces pore coalescence, but these differences are very small. The overall densification was therefore evaluated using relative density and dilatometry data, confirming nearly full densification of the samples. These microstructures are relatively similar, in agreement with the dilatometry results, indicating good densification for all samples without open porosity. The darker zones typically indicate the silica-rich phases [23].

**(A) 2 K/min**

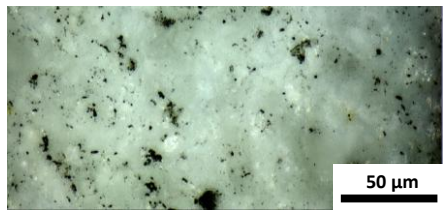

**(B) 5 K/min**

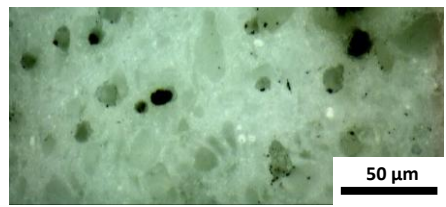

**(C) 10 K/min**

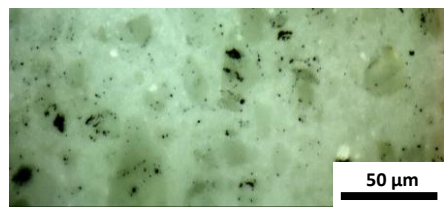

*Figure 3. Optical microscopy of sintered porcelain at different heating rates.*

## 3.2. Activation Energy Determination

To quantify the thermally activated nature of the sintering process, the activation energy was estimated using two complementary methods: the Master Sintering Curve (MSC) [38] and the Wang-Raj (WR) approach [39], as shown in Figure 4. In Figure 4A, the densification curve are plotted versus temperature and show the expected shift to lower temperatures for the lower heating rates.

The MSC method combines the densification data from different heating rates into a unified densification curve applying on the sintering work integral. An error minimization approach is then used to assess the sintering activation energy. As reported in the inset in Figure 4B, the minimization yields an apparent activation energy of 350 kJ/mol, corresponding to the best superposition of the densification curves. This value is consistent with the 300 kJ/mol value obtained in ref [23] and the 200-400 kJ/mol range of ref [44].

To validate this result, the WR method was applied, which uses a linear regression with respect to inverse temperature to assess the activation energy. Linear fitting of the WR plot (Figure 4C) provides an independent estimate of the activation energy, yielding a value of 357 kJ/mol. The close agreement between MSC and WR values reinforces the consistency of the experimental data and confirms the thermally activated densification mechanism.

These activation energies falls within the typical range reported for viscous sintering of porcelain-based ceramics, and they will be used for the calibration of the finite element model in the next section.

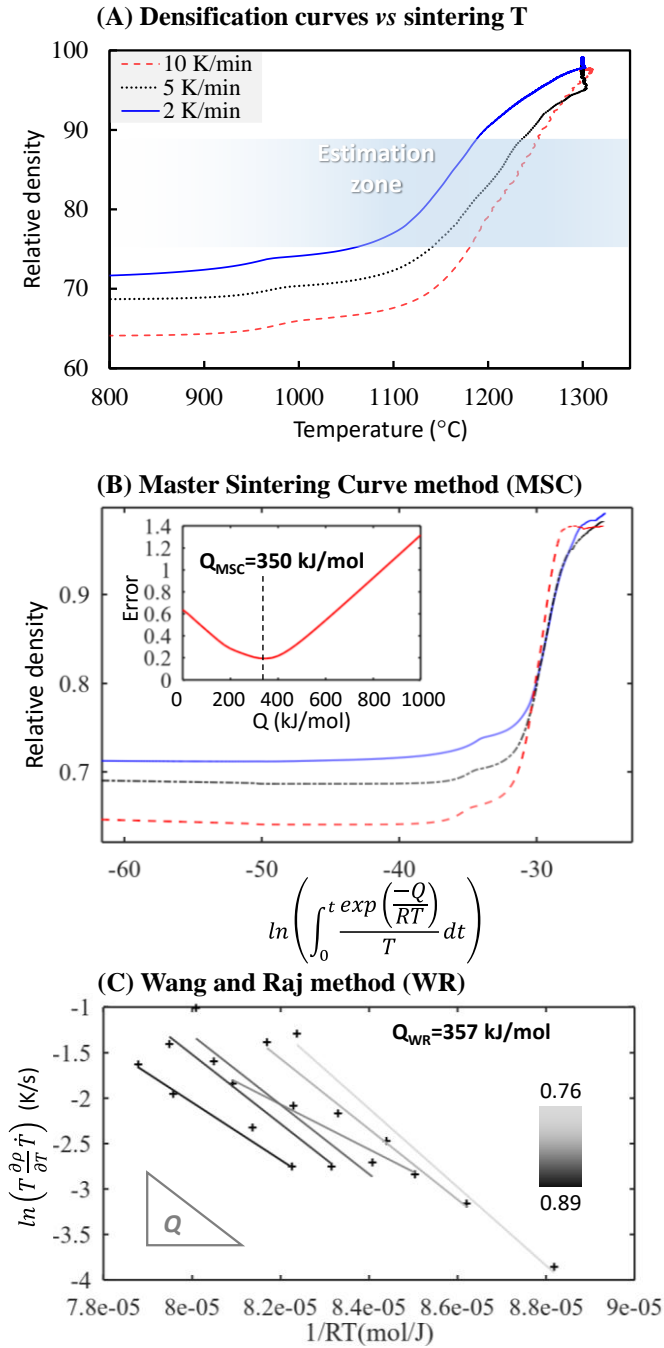

*Figure 4. Activation energy determination from densification data. A) Relative density curves at different heating rates. (B) Master Sintering Curve (MSC) analysis using time-temperature integration. (C) Wang-Raj (WR) method based on the logarithmic derivative of densification rate.*

### 3.3. Assessment of bulk modulus by regression-based minimization

To implement the sintering model in the finite element framework, the apparent viscosity constant ($\eta_0$) and the bulk modulus (3) expression need to be identified using the regression equation (2). The slope of this regression equation gives an activation energy that is dependent on the bulk modulus expression. Consequently, if the bulk modulus function does not correspond to the WR (or MSC) real activation energy (which is independent on porosity function by multiple heating rates) the apparent slope will be different. A minimization study can then be conducted to assess the bulk modulus exponent so it reproduce the identified activation energy. This regression plot is reported in Figure 5A. The slope is givin J/mol and the $R^2$ value is used to indicate the accuracy of the model in intermediate sintering. The Bulk modulus exponent are tuned so it reproduce the WR activation energy (minimizing the error $|Q-Q_{WR}|$) and the with a $R^2$ closer to 1. The critical porosity values where fixed to 0.35 for the initial value (5% margin from initial porosity) and the final critical porosity was set to 0.01 (1%) corresponding to the final porosity of the lowest experimental porosity obtained. The exploration study is reported in Figure 5B. There are different bulk modulus exponents compatible with the WR activation energy, but the one having good accuracy correspond the following bulk modulus expression.

$$\psi = \frac{2}{3} \frac{0.35 - \theta}{(\theta - 0.01)^{1.5}} \tag{10}$$

The best fit was obtained for $\zeta = 1.5$ and $\gamma = 1.0$ corresponding to a high degree of correlation ($R^2 > 0.98$). These values define the porosity-dependent part of the model and the constant term of the viscosity is taken from the origin of the regression equation. After taking average particle radius of 3.48 µm (measured) and the surface energy of mullite (0.66 J/m² from [45], the identified parameters can be used in the COMSOL model to simulate shrinkage and porosity evolution.

The finite element simulations were performed in COMSOL Multiphysics®. From orevious mesh analysis [46], second-order mesh elements were used and the absence of intermediate nodes in bar thickness were avoided.

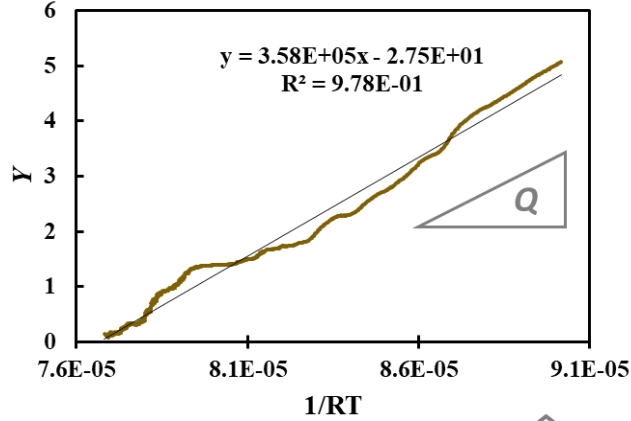
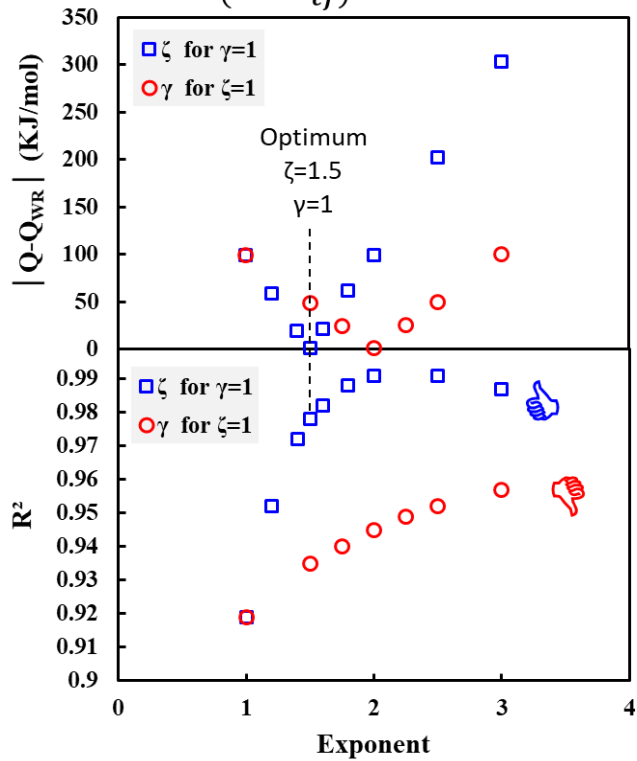

Figure 5. Identification of viscosity model parameters. (A) Regression of experimental data using the modified sintering function. (B) Analysis of the influence of Skorohod's exponents ($\gamma$ and $\zeta$) on model fitting accuracy for the regression-minimization approach.

### 3.4. Sintering Anisotropy Assessment

To evaluate the accuracy of the calibrated sintering model, the porosity evolution was first analytically simulated using Equation (4). The results are shown in Figure 6A. In the initial stage, the behavior (at 1000°C) is not considered, because it is far from the temperature at which deformation can occur. Indeed, the primary objective of the study is to be able to predict the sintering distortions of the 3D printed parts. Aside from the initial stage, the porosity elimination curve is well reproduced in the intermediate and final stage.

The finite element simulation of the 2 K/min dilatometry pellet sintering is reported in Figure 6B. Initially, the input model is isotropic. The anisotropy is added by an anisotropic sintering stress porosity function (*Plr, Plz*). The layer-by-layer deposition of fused filament fabrication induces microstructural anisotropy and preferential pore orientation. Olevsky demonstrates [41] that elliptical pores (like those that can result from printing [47]) can imply anisotropic sintering stresses (*Plr, Plz*) and resulting anisotropic shrinkage. Sarbandi et al. [48] model the anisotropy in the same way, pointing out the computational stability of such an approach.

It is acknowledged that 3D-printed green bodies exhibit interlayer porosity and non-uniform binder distribution that lead to local heterogeneity. In the present continuum framework, porosity is treated as an effective averaged variable describing the global densification behavior. The introduced anisotropic sintering stress functions (*Plr, Plz*) implicitly account for the macroscopic effects of such microstructural heterogeneities, allowing reliable prediction of shrinkage and distortion at the part scale.

The set of functions reported in inset in Figure 6B have been obtained and succeed in reproducing the experimental shrinkage reported by the dilatometer. The FEM simulation use a numerical probe located by the red point to record the axial (Z) and radial (R) shrinkages in the simulation (exactly like in the experiment). At this stage, the anisotropic sintering model is fully identified.

The deviation between simulated and experimental shrinkage remains below 5% for both radial and axial directions, while the predicted porosity evolution differs by less than 3% from the measured data. These small discrepancies in the beginning of the sintering are mainly attributed to initial liquid phase and rearranging effects always complex to simulate and to local variations in the green density. Overall, the model reproduces the experimental results with high fidelity, confirming its predictive accuracy for porcelain sintering.

The next step consists of adjusting the creep part by the shear viscosity.

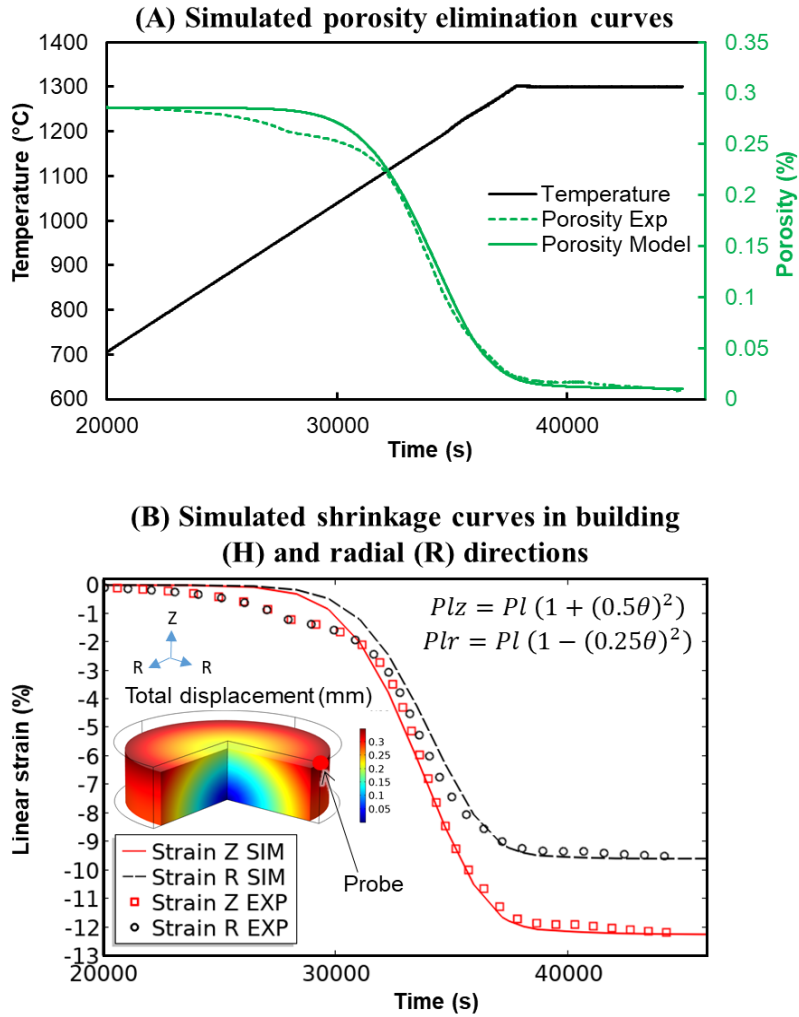

*Figure 6. Validation of porosity and shrinkage predictions. (A) Comparison between experimental and simulated porosity evolution. (B) Comparison of anisotropic shrinkage in vertical (Z) and radial (R) directions.*

*3.5. Shear Viscosity Assessment and Model Validation*

The assessment of the shear viscosity is mandatory to have a reliable model both in densification and to capture its deformation sensitivity. Indeed, the gravitational forces and friction with the support can develop stresses that can lead to deformations. The latter are particularly important for thin wall structures with insufficient support. The identified sintering model already captures the viscosity temperature dependence but the shear modulus still needs to be evaluated. From equation (9) the shear modulus is explored. The porosity function assumes Skorohod's expression [41] with an exponent of 2. A correcting factor ($\beta$) is adjusted; the exponent can be modified if the correcting factor is not sufficient. A thin 3D printed empty cylinder is sintered on the side to exaggerate the sintering distortions. In this test, the distortion is high and it is used to adjust the shear viscosity factor ($\beta$). The result is reported in Figure 7A, The graph shows the H/R simulated ratio for different shear viscosity correction factors. A factor of 8.7 is need to reproduce the 0.83 H/R experimental ratio. The simulated deformation closely reproduces the experimentally observed distortion with in particular the effect of the support on the lower part.

In a second validation case, a bar-shaped sample was sintered in a horizontal position to induce gravitational deflection. Figure 7B compares the final displacement of the printed bar to the simulated result. This simulation use the identified 8.7 correcting factor of the shear viscosity. The model predicts both the curvature and the amplitude of the deflection with good accuracy, confirming the suitability of the viscosity law and calibrated parameters under non-symmetric loading conditions.

These validations on real geometries confirm the applicability of the model beyond simple test cases and highlight its potential for predictive simulation of complex ceramic parts processed via additive manufacturing. This also demonstrates the viscosity identified from sintering dilatometry is slightly different from the one responsible for the deformation. Indeed the deformation is mainly due to shear contribution while sintering is due to liquid phase particle rearrangement and dissolution precipitation mechanisms. The development of mullite needles network [43] at the final stage (in non-densifying zones) can explain why the shear deformation sensitivity is different from the apparent viscosity of sintering which is mainly governed by the liquid phase development. For traditional solid state sintering similar differences are observed, Desplanques *et al* [49] proposed a grain boundary sliding mechanistic transition based on threshold stress to model this kind of deformation phenomenon for ceramics like alumina.

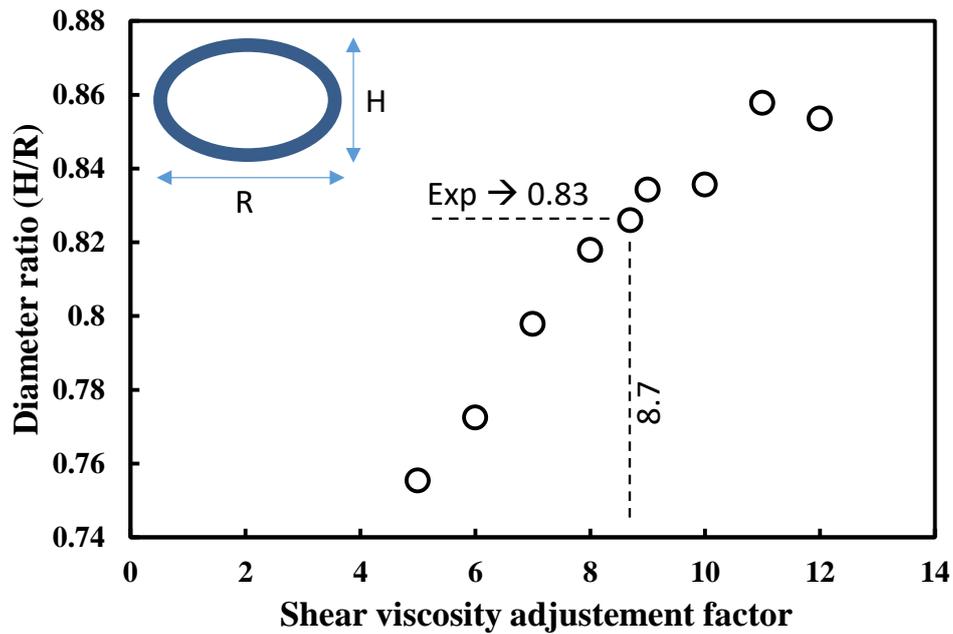

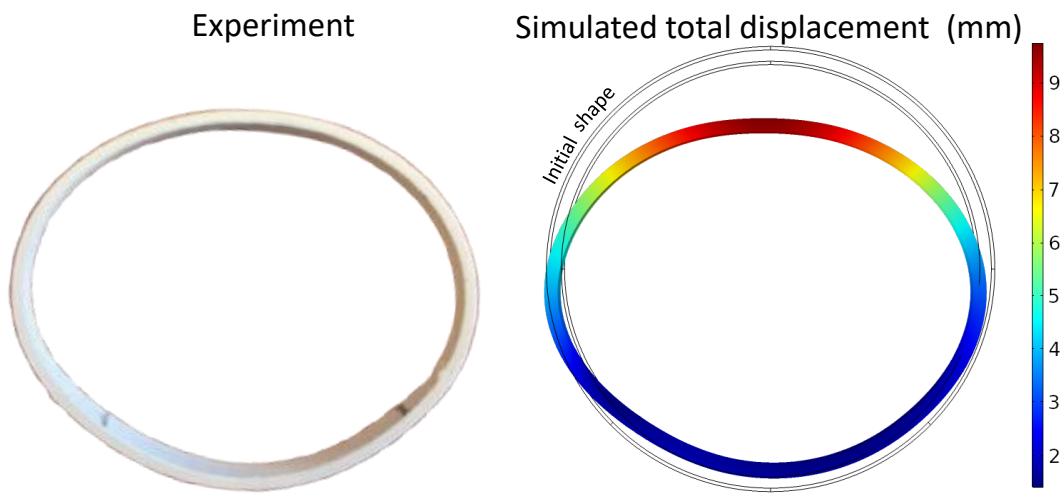

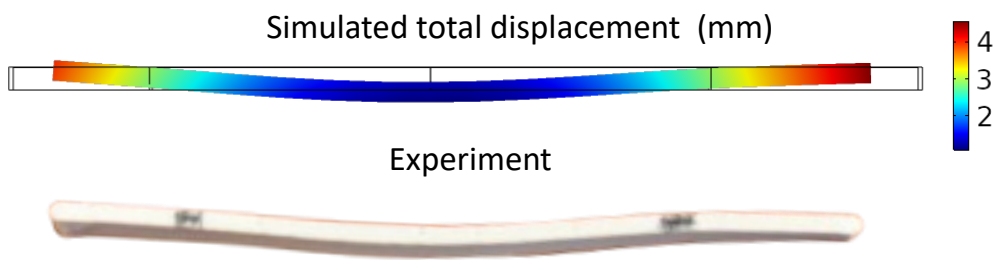

*Figure 7. Validation of the sintering model on real printed geometries. (A) Ring slumping test: comparison between experimental and simulated deformation. (B) Bar deflection test: verification of sintering-induced bending due to gravity.*

These results demonstrate that this identification method of the sintering model accurately reproduces both densification and deformation behavior across a range of conditions, from simple geometries to realistic printed parts. By integrating experimental data with numerical simulation, the model offers a reliable tool for predicting shrinkage, porosity evolution, and shape distortion during porcelain sintering. This combined approach supports future efforts to optimize process parameters, reduce trial-and-error in ceramic manufacturing, and enable the design of more complex components with improved dimensional accuracy.

## 4. Conclusion

This work introduced a step-by-step methodology combining anisotropic sintering dilatometry and finite element calibration to predict the deformation of 3D-printed porcelain parts. Porcelain components produced using Zetamix by Nanoe® filaments were debinded and prepared for sintering analysis. Dilatometry tests at 2, 5, and 10 K/min were conducted to characterize the sintering kinetics and the anisotropy originating from the printing process. The sintering model was identified through a porosity-independent evaluation of the activation energy using the Master Sintering Curve and Wang–Raj methods, while the bulk modulus exponent was determined by minimization and regression. Once the densification model is identified, the shrinkage anisotropy is adjusted in the FEM using the experimental geometric shrinkage curves of the pellets. Finally, the shear viscosity was tuned using the sintered ring geometry to correctly capture the deformation behavior. The resulting model was tested on overhang bar shapes and successfully reproduced the observed deformation trends.

**Funding**

This work has been funded by BPI France project "ZETAFACTORY" number "259872"